%% file: DeadLayerPaper4_29_13_arXiv.tex
\documentclass[1p,number,sort&compress]{elsarticle}
\usepackage{lineno}

 \usepackage{graphicx}

\usepackage{amssymb}
\usepackage{pifont}
\usepackage{xmpmulti}
\usepackage{amsmath}
\usepackage[lofdepth,lotdepth]{subfig}
\usepackage{color}

\newcommand{\pin}{\textit{p-i-n}}

\newcommand{\tref}[1]{Table~\ref{#1}}
\newcommand{\fref}[1]{Fig.~\ref{#1}}

\begin{document}

\begin{frontmatter}
\input{Author_DeadL_4_29_13_bw}
\input{Abstract_DeadL_4_27_13_hr}

\end{frontmatter}
\input{Introduction_DeadL_4_29_13_arXiv}
\input{MaterialMethods_DeadL_4_27_13_arXiv}

\input{ResultsDiscussion_DeadL_4_29_13_arXiv}
\input{Conclusion_DeadL_4_29_13_arXiv} 
\input{ack}

\bibliographystyle{unsrtnat}
\bibliography{BibDeadL}

\end{document}

%% file: Author_DeadL_4_29_13_bw.tex


\title{Dead layer on silicon p-i-n diode charged-particle detectors \\ 
}

\author[uw]{B.\,L.~Wall\corref{corauths}}
\author[uw]{J.\,F.~Amsbaugh}
\author[kitipe]{A.~Beglarian}
\author[kitipe]{T.~Bergmann}
\author[uw]{H.\,C.~Bichsel}
\author[uw]{L.\,I.~Bodine}
\author[uw]{N.\,M.~Boyd}
\author[uw]{T.\,H.~Burritt}
\author[setif]{Z.~Chaoui}
\author[unc,tunl]{T.\,J.~Corona}
\author[uw]{P.\,J.~Doe}
\author[uw]{S.~Enomoto}
\author[kitik]{F.~Harms}
\author[uw]{G.\,C.~Harper}
\author[unc]{M.\,A.~Howe}
\author[uw]{E.\,L.~Martin}
\author[uw]{D.\,S.~Parno}
\author[uw]{D.\,A.~Peterson}
\author[kitipe]{L.~Petzold}
\author[kitiekp]{P.~Renschler}
\author[uw]{R.\,G.\,H.~Robertson\corref{corauths}}
\author[kitik]{J.~Schwarz}
\author[kitik]{M. ~Steidl}
\author[uw]{T.\,D.~Van Wechel}
\author[pnnl]{B.\,A.~VanDevender\corref{corauths}}
\author[kitipe]{S.~W\"{u}stling}
\author[uw,unc,tunl]{K.\,J.~Wierman}
\author[unc,tunl]{\\ and J.\,F.~Wilkerson}

\address[kitik]{Institute for Nuclear Physics, Karlsruhe Institute of Technology, Karlsruhe, Germany}
\address[kitipe]{Institute for Data Processing and Electronics, Karlsruhe Institute of Technology, Karlsruhe, Germany}
\address[kitiekp]{Institute for Experimental Nuclear Physics, Karlsruhe Institute of Technology, Karlsruhe, Germany}
\address[pnnl]{Pacific Northwest National Laboratory, Richland, WA, USA} 
\address[unc]{Dept.\,of Physics and Astronomy, University of North Carolina, Chapel Hill, NC, USA}
\address[tunl]{Triangle Universities Nuclear Laboratory, Durham, NC, USA}
\address[uw]{Center for Experimental Nuclear Physics and Astrophysics, and Department of Physics, University of Washington, Seattle, WA, USA} 
\address[setif]{Laboratory of Optoelectronics and Devices, Faculty of Science,
University of Setif, Algeria}


\cortext[corauths]{Corresponding authors.  wallbl@uw.edu, rghr@uw.edu, brent.vandevender@pnnl.gov}



%% file: Abstract_DeadL_4_27_13_hr.tex
\begin{abstract}
Semiconductor detectors in general have a dead layer at their surfaces that is either a result of natural or induced passivation, or is formed during the process of making a contact.  Charged particles passing through this region produce ionization that is incompletely collected and recorded, which leads to departures from the ideal in both energy deposition and resolution.  The silicon \pin{} diode used in the KATRIN neutrino-mass experiment has such a dead layer.  We have constructed  a detailed Monte Carlo model for the passage of electrons from vacuum into a silicon detector, and compared the measured energy spectra to the predicted ones for a range of energies from 12 to 20 keV. The comparison provides experimental evidence that  a substantial fraction of the ionization produced in the ``dead'' layer evidently escapes by diffusion, with 46\% being collected in the depletion zone and the balance being neutralized at the contact or by bulk recombination.   The most elementary model of a thinner dead layer from which no charge is collected is strongly disfavored.
\end{abstract}

%% file: Introduction_DeadL_4_29_13_arXiv.tex
\section{Introduction}\label{sec:intro}

Semiconductor detectors have surface ``dead'' layers in which ionizing particles can deposit energy, creating charge that is, at best, incompletely collected by the readout electronics~\cite{Knoll,Leo}.  These layers correspond to the doped regions defining the semiconductor properties of the device, ohmic contacts to bias voltage and readout electronics and/or undepleted volumes of under-biased detectors.  Dead-layer effects have received detailed attention in the spectroscopy of low-energy gamma and X-rays, which deposit energy in or near the dead layer.   Hartmann {\em et al.} \cite{Hartmann1996191} carried out measurements of the charge collected from dead layers in $pn$-junction detectors.  A study of the charge collection from Li-diffused contacts on Ge detectors has been reported by Aguayo {\em et al.} \cite{Aguayo2013176}.   Similarly, for charged particles, energy loss in even the thinnest of dead layers can amount to a significant fraction of the total energy deposition for radiations of interest in nuclear and particle physics experiments.  A 20-keV electron, for instance, will deposit approximately 1\,\% of its total energy in a 100-nm silicon dead layer.  In high-precision measurements, the thickness of the dead layer and the fate of charges created therein must be taken into account.  In the work that is the subject of this paper,  it is found that almost half of the charge created in a silicon \pin{} diode dead layer is transported by diffusion into the depleted region where it  contributes to signal formation by drifting and being collected at the active contacts.  

\begin{figure}[tb]
\begin{center}
\resizebox{0.5\textwidth}{!}{\includegraphics{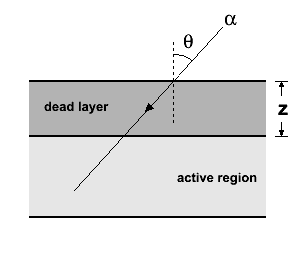}}
\caption{\label{fig:angleMethGeom} The geometry of the alpha method.}
\end{center}
\end{figure}

The standard method~\cite{IEEE300,Knoll} for determining dead-layer thickness uses an alpha source.  The deficit in recorded energy due to uncollected charge created along the path length through the dead layer is measured as a function of angle of incidence $\theta$ (see \fref{fig:angleMethGeom}).
If it is assumed that no charge is collected from the dead layer, the measured energy $E$ deposited in the active region of the detector is approximately:
\begin{eqnarray}
E(\theta) &=& E_0 -  \Delta E  \left( \frac{1}{\cos{\theta}} \right)  \label{eqn:angle_method},
\end{eqnarray}
where $\Delta E$ is the energy loss in the dead layer evaluated at the incident energy $E_0$ and normal incidence.  The IEEE Standard~\cite{IEEE300} calls only for $\Delta E$ to be reported, but it is common practice to convert it to a thickness by using the stopping power:
\begin{eqnarray}
z &=& \Delta E \left< \frac{dE}{dx} \right>^{-1} \label{eqn:stoppingpower},
\end{eqnarray}
where $\left< dE/dx \right>$ is the  stopping power evaluated at the incident energy $E_0$, and $z$ is the dead-layer thickness.   As we show, this step can lead to serious underestimates of the thickness when there is partial charge collection from the dead layer.   The  textbook model of a dead layer from which no charge is collected underestimates the thickness of the layer by about a factor of two.


In the KATRIN neutrino-mass experiment~\cite{katrin_dr2004}, electrons that pass through two energy-selective spectrometers are detected in a 148-pixel silicon \pin{} diode detector in the spectrometer focal plane.  Precision and stability are essential features of the detection system to control systematic errors.  Time-dependent changes in the thickness of \pin{} diode dead layers have been reported~\cite{Lising:2000pa}, and an {\em in situ} method for measuring the dead layer is therefore highly desirable.   The complex magnetic, electrostatic, cryogenic, and ultra-high-vacuum environment of the KATRIN focal-plane detector (FPD) makes the standard alpha-particle angular-variation method impractical.  A new approach that makes use of monoenergetic electrons normally incident on the detector surface has been developed in response to this need.   A related objective was a comprehensive physics-based description of the response function for electrons.  An accurate description of the line shape allows the signal to be extracted by maximum likelihood, which optimizes the statistical precision.  To meet both these objectives, the KATRIN Electron Scattering in Silicon (KESS) Monte Carlo code has been developed (see \cite{Prall:2012rx,Renschler2012493,Renschler_thesis} and references therein).  Not only has this approach met the original objectives, it has revealed unexpectedly detailed information about the properties of \pin{} diode dead layers.

Before the development of KESS, an initial experimental investigation~\cite{IEEE4178978} of the dead-layer measurement problem explored two paths.  One approach is to apply the standard angular-variation method to monoenergetic electrons incident on the detector surface.   The second approach, which does not require angular variation,  makes use of the variation of energy lost in the dead layer as a function of the incident energy.  In that scenario, the entrance angle is always normal but the energy of the electrons is varied.  The energy dependence of the stopping power permits extraction of the dead-layer thickness from a fit to data recorded at various incident electron energies.  Electrons of useful energies can be created artificially with electron guns or naturally with Auger and conversion-electron sources.  The latter type of source  has been used to measure silicon-detector dead layers~\cite{Petersen1984582}.      In our work~\cite{IEEE4178978} an electron gun was used to study  a 64-pixel \pin{} diode using both methods.  The data were analyzed using a simple continuous-slowing-down approximation (CSDA).  The energy loss was described by a Bethe-Bloch function~\cite{Leo}, and the observed energy was reproduced by adjusting the dead-layer thickness to produce the best fit.   While the methods were sensitive enough to yield quite precise values, they disagreed on the value for the dead-layer thickness by 10\,\%.  Since CSDA was known to be  particularly inappropriate for electrons in any case~\cite{PhysRevA.16.1061}, work began on the Monte Carlo transport code KESS.  In KESS, electron transport is treated microscopically on a collision-by-collision basis, accounting for energy loss and angular deflection in each individual interaction, and tracking  primary and scattered electrons until their energies fall below preset precision levels.

The widths and low-energy tails of monoenergetic electron energy spectra are, fractionally, much more sensitive to energy loss in the dead layer than are the mean and most probable energies, and are also less sensitive to the electronic stability.  With KESS, a detailed analysis of peak shapes became possible and was applied to electron-source measurements made on the KATRIN FPD device provided by Canberra.  It was never possible to achieve good fits in the textbook model of a detector in which the dead layer is a slab of completely inert material adjoined to a slab of fully depleted active material.   In contrast,  good fits could be obtained in a diffusion model with a dead layer from which approximately half the deposited ionization was recovered into the active region.  

In the following sections, we detail the apparatus and analysis from which the above conclusions have been drawn.     Section~\ref{sec:methods} describes the KATRIN FPD, and collection and analysis of data.  Section~\ref{sec:results} compares KESS Monte Carlo fits in the textbook and diffusion models and demonstrates the improved fidelity of the diffusion model.

%% file: MaterialMethods_DeadL_4_27_13_arXiv.tex
\section{Materials and Methods}\label{sec:methods}

\begin{figure}[h]
 \centering
 \includegraphics[width=4in]{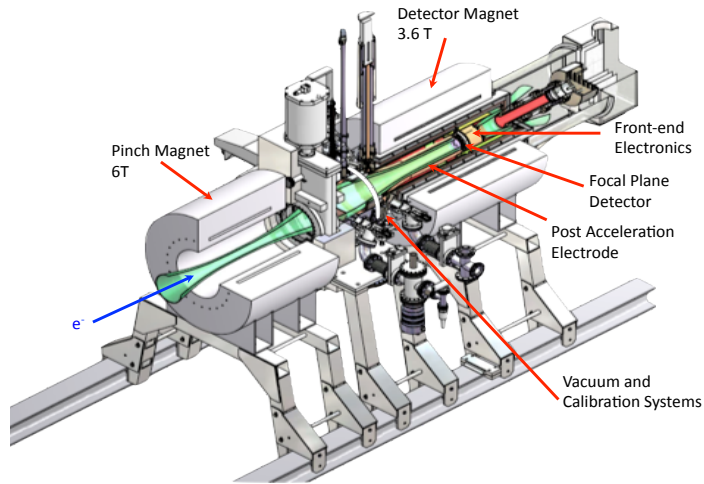}
 \caption[FPDsystem]{A cutaway view of the KATRIN focal-plane detector system showing the major components.  The ``flux tube'' shown in green along the axis is the bundle of electron trajectories that are transported to the detector.}
 \label{fig:FPDsystem}
 \end{figure}

\begin{figure}[h] 
     \centering
    \subfloat[][]{
     \includegraphics[width=2.37in]{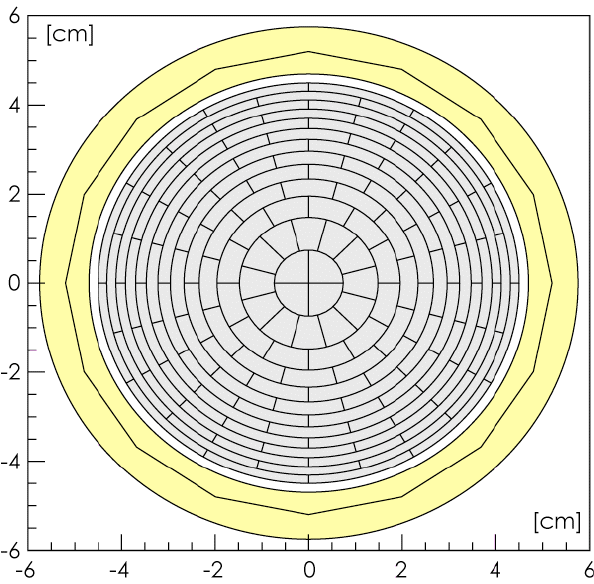} 
     \label{fig:PixelMask}
     }
     \qquad
      \subfloat[][]{
     \includegraphics[width=2.3in]{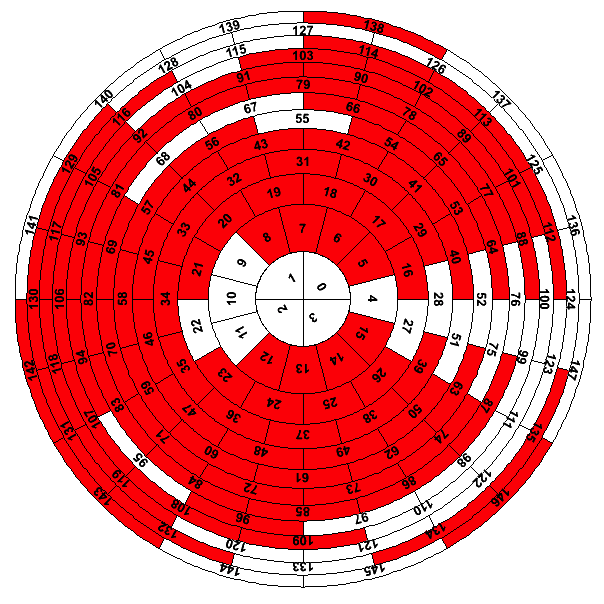}
     \label{fig:PixelNumber}
     }
     \caption[Focal Plane Detector Layout]{ \subref{fig:PixelMask} The pixel segmentation on the junction or contact side of the focal-plane detector. \subref{fig:PixelNumber} Numbered pixel map, with colored pixels showing the ones included in the analysis. The white pixels were removed from the analysis due to poor electron-source illumination, missing energy calibration, baseline oscillations, high crosstalk between channels, low-energy noise bursts, and dead preamplifier channels.}
     \label{fig:FPDLayout}
  \end{figure}

  The KATRIN detector system was first assembled at the University of Washington.  The data used to make the dead-layer measurements on the FPD were taken in the spring of 2011.  A view of the detector system is shown in \fref{fig:FPDsystem}. 
    The FPD~\cite{VanDevender:2012rx} is a 500-$\mu$m-thick custom monolithic silicon \pin{} diode array manufactured by Canberra.  Particles enter through an unsegmented ohmic {\em n}++ implanted face, and junctions are formed on the other side at {\em p}+ implanted pixels that have a TiN overlayer.  The detector has a total active area of 65.3\,cm$^2$, shared equally by 148 44.1-mm$^2$ pixels.    Figure~\ref{fig:PixelMask} shows the pixel segmentation for the focal-plane detector.  
 Each pixel has a capacitance of 8.2\,pF and is instrumented with a charge-integrating preamplifier followed by a variable-gain amplifier stage.  The signal path is optically coupled to the KATRIN data-acquisition crate, which processes each channel's signal through a series of bandpass filters and a programmable amplifier leading to serial 20-Mhz 12-bit ADCs.   Two cascaded programmable trapezoidal filters determine event energy and time.  The ORCA real-time data-acquisition program~\cite{ORCA_FPGA} performs crate data readout. 
  The detector section is equipped with a photoelectron source, a titanium disc held at a potential of 0 to $-20$\,kV and illuminated by a UV diode.  The negative potential applied to the disc accelerates the photoelectrons to the entrance surface of the detector, which is biased at $+120$\,V with respect to ground.   A 3.6-T magnetic field images the photoelectrons from the disc onto the FPD.   The combination of the magnetic field and the potential on the disc assures that any  electrons backscattered from the detector are reflected to the same pixel, with rare exceptions for strikes near a pixel boundary.   Some electrons make several transits through the dead layer before being absorbed.  This feature provides additional sensitivity  to the thickness of the dead layer in the analysis.   

  Rather than using an approximation for the energy loss in the dead layer, our method uses the entirety of the energy spectrum produced by monoenergetic electrons to evaluate the dead-layer thickness~\cite{Wall_thesis}.  Simulated spectra, generated by KESS~\cite{Renschler2012493,Renschler_thesis}, are compared to electron-source spectra using a $\chi^2$-minimization technique.	Electron-source data were measured for eight different voltage settings from 12.6\,kV to 19.6\,kV in 1-kV increments.  Voltage settings lower than 12.6\,kV were not used because the low-energy tails of those spectra were cut off by the electronics hardware threshold, which varied by channel in the range of 5 to 8\,keV.  An $^{241}$Am source provided an independent energy calibration from two gamma peaks at 26.3448 keV and 59.5412 keV.

	Each pixel is modeled as an isolated single \pin{} diode detector illuminated with  monoenergetic electrons.  The simulated detector has an active layer and a dead layer of silicon.  For each electron energy setting, simulated spectra were generated for dead-layer thicknesses of 50\,nm to 500\,nm in 10-nm increments.  Two models of the dead layer were used: the elementary inert-slab model where energy deposited in the dead layer is lost, and a  diffusion model.  The diffusion model used the same simulation as the inert-slab model, but a fraction $d$ of the energy deposited in the dead layer was added to the energy deposited in the active region.   
  
  	Electrons were simulated at energies matching the source data.  The effect of the magnetic field and the source-disc potential on backscattered electrons was modeled by immediately negating the momenta of electrons exiting the detector to return them to the detector.  This is a reasonable approximation as the transit time for a backscattered electron to return to the detector is of order nanoseconds, well within the 6.4-$\mu$s shaping time of the data-acquisition system.
\begin{figure}[ht] 
     \centering
     \subfloat[Before Noise Convolution][]{
            \includegraphics[width=2.3in]{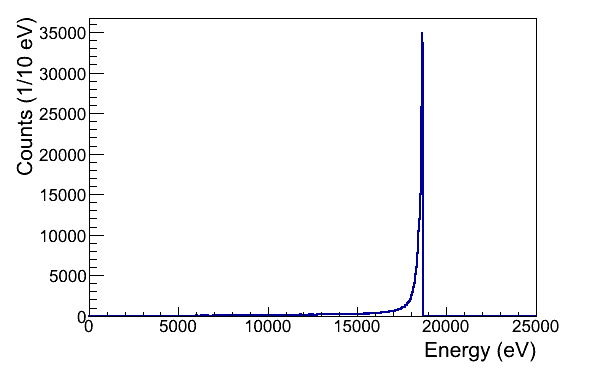} 
            \label{fig:simulatedSpectra}
                  }
\qquad
      \subfloat[After Noise Convolution][]{
       \label{fig:simulatedSpectraConvolved}
       \includegraphics[width=2.3in]{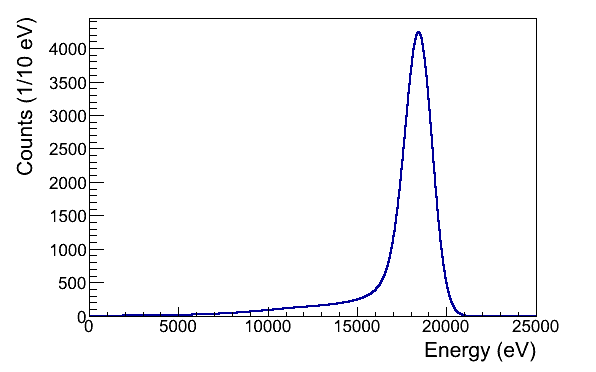}
        }\\

     \caption[Simulated Electron Energy Spectra]{  \subref{fig:simulatedSpectra} Simulated 18.6-keV electron energy spectrum with a 3.6-T magnetic field in a \pin{} diode with a 150-nm dead layer.  \subref{fig:simulatedSpectraConvolved}  The simulated spectrum convolved with an electronic-noise Gaussian and a Fano-noise Gaussian.
    }
  \end{figure}
Figure~\ref{fig:simulatedSpectra} is a simulated electron spectrum generated by KESS for a 150-nm dead layer.  Electronic noise is treated by convolving the KESS spectrum with a Gaussian whose width is determined by electronic calibration data (typically 640-eV standard deviation).  Fano noise is added by convolution with a Gaussian of standard deviation 
    \begin{eqnarray} 
       \sigma_{{\rm Fano}}&=&\sqrt{EF\epsilon},
           \end{eqnarray}
    where $E$ is the energy deposition, $F=0.143$ is the Fano factor, and $\epsilon=3.62$\,eV is the mean energy to create an electron-hole pair in silicon~\cite{Knoll}.   These two convolutions transform the spectrum in \fref{fig:simulatedSpectra}  into the one in \fref{fig:simulatedSpectraConvolved}.  

     Each measured electron-source spectrum is compared to a library of spectra simulated for different dead-layer thicknesses.  Before comparing simulation to data, the number of events in the simulated spectrum is normalized to the number in the electron-source spectrum in the comparison region.  The energy of the simulation peak is brought into alignment with that of the source data by a free offset parameter.  The comparison results immediately in a $\chi^2$ value for each dead-layer thickness tested.  Apart from the energy offset, dead-layer thickness, and charge-collection fraction there are no other free parameters.

  The dead-layer thickness is independent of electron energy, so multiple electron energies can be used to increase the data set and better constrain the fit.  A total $\chi^2$ at each dead-layer thickness is evaluated from the sum of the $\chi^2$ values from each energy setting.  The best-fit value for the dead-layer thickness is the minimum of a quadratic fit to total $\chi^2$ as a function of the dead-layer thickness.  Marginalizing over other fit parameters, the $\chi^2$ minimum is increased by one to determine the statistical error.

%% file: ResultsDiscussion_DeadL_4_29_13_arXiv.tex
\section{Results and Discussion}\label{sec:results}
\begin{figure}[ht] 
     \centering
     \includegraphics[width=5.0in]{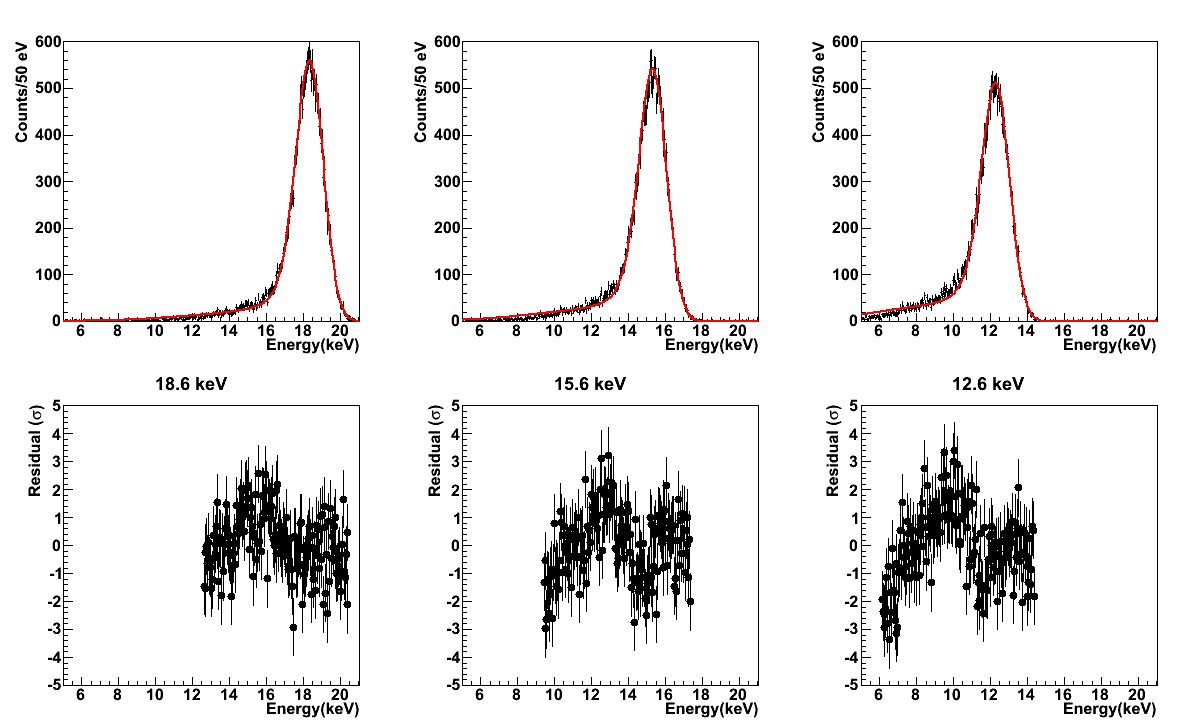} 
     \caption[Electron Source Spectra Fits and Residuals for Pixel 31]{(Top row) Pixel-31 electron-source data in black fit with the simulation without diffusion and a dead layer of 110.0\,nm in red for electron-source energies of 18.6 ($\chi^2/{\rm n.d.f.}=1.04$), 15.6 ($\chi^2/{\rm n.d.f.}=1.48$) and 12.6 ($\chi^2/{\rm n.d.f.}=2.16$) keV.  The lower three plots are residuals normalized to the standard deviation of the electron-source data.  The residuals display only the points within the fit window.}
     \label{fig:Pix31Fit_Field_noDiffusion}
  \end{figure}

  Evaluating the dead-layer thickness with a simulation using an inert-slab dead layer without charge diffusion leads to poor fits of the data.  An examination of the fits for pixel 31 in \fref{fig:Pix31Fit_Field_noDiffusion} shows that the simulated spectra do not correctly fit the data in the low-energy tails.  The effect is nearly unnoticeable in the 18.6-keV spectrum as the $\chi^2/{\rm n.d.f.}$ is $1.04$, but as the incident electron energy decreases the discrepancy increases.  For the 12.6-keV spectrum, the simulation overestimates the data below 8\,keV, and underestimates the data from 8\,keV to 11\,keV.  The discrepancy appears in the residual as the steep linear slope from 7\,keV to 11\,keV.  
  
  Changing to the alternate model with diffusion improves the fit of the simulated spectra to the electron-source spectra. Such a model is entirely consistent with the microscopic picture of a dead layer consisting of heavily doped silicon.  Negligible drift field is present there, and ionization escapes only by diffusion or recombination.  A charge promoted into the conduction band by ionizing radiation will typically live for more than $10^{-5}$\,s before it recombines.  In a time $t$, charges created at a point will diffuse isotropically a typical distance 
\begin{eqnarray}
\sigma &=& \sqrt{2 \mu \frac{kT}{e} t}\label{eqn:sigma},
\end{eqnarray}
where $\mu$ is the charge-carrier mobility, $k$ is the Boltzmann constant and $T$ is the temperature~\cite{Knoll}.  The time required to diffuse 100\,nm, a typical dead-layer thickness, is measured in picoseconds.  The diffusion time is orders of magnitude shorter than the recombination time.  Mobile charge carriers generated at random points in a uniform, field-free region between two boundaries will on average diffuse to either boundary with equal probability.  Therefore, approximately half the charges will diffuse to the active region, where they are collected, and half to the contact, where they are neutralized.   It is feasible to model the doping profile, internal fields, and charge-collection efficiency for a detector accurately~\cite{Hartmann1996191}.  However, although doping profiles can be complex, simplification arises when there exists a boundary between the low-field region where charge carrier motion is dominated by diffusion (or recombination) and the high-field region dominated by drift.  There are in that case only two relevant parameters, the depth of the boundary, which is defined to be the dead-layer thickness, and the charge-collection efficiency for charge carriers in the dead layer.  The situation, the usual one for ion-implanted charged-particle detectors, is illustrated in Fig.~1a in~\cite{Hartmann1996191}.

The best match between data and simulation is found when, as expected, almost half of the dead-layer charge is added to the active-region charge.  The dead layer that best matches simulation to data is much thicker in the diffusion model than that which optimizes the match in the inert-slab model.

\begin{figure}[ht] 
     \centering
     \includegraphics[width=5.0in]{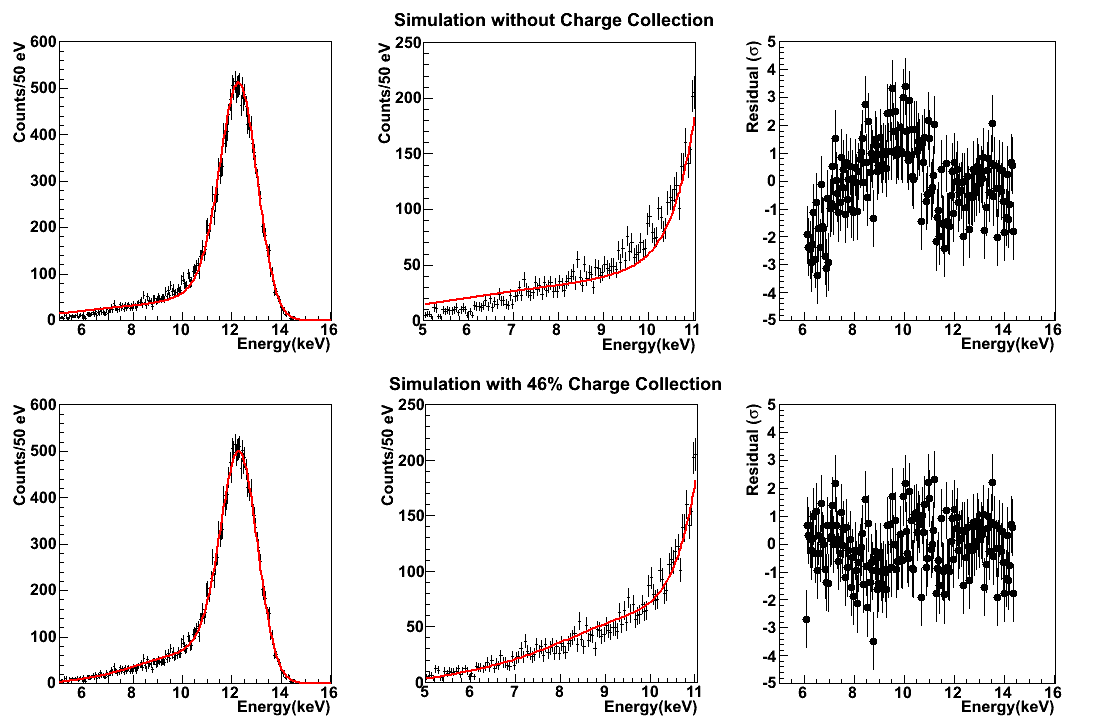} 
     \caption[]{Pixel-31 12.6-keV electron-source data and the best-fit simulation in red. The top row shows simulation data without a charge-collection fraction and a 110.0-nm dead layer, while in the bottom row the simulation is for a charge-collection fraction of 0.46 and a dead-layer thickness of 160 nm.  The middle panels are expanded views of the leftmost panels for the 5- to 11-keV region.  The rightmost panels are residuals normalized to the error $\sigma$ of the electron-source data.  The residuals only display the points within the fit window.}
     \label{fig:Pix31Fit_Field_Diffusion}
  \end{figure}

 Assuming the charge-collection fraction is the same for each pixel, a manual search over a range of values of $d$ disclosed a  minimum $\chi^2$  at a charge-collection fraction $d=45.9 \pm 0.1$ (stat.) percent ($\chi^2/{\rm n.d.f.}=1.15)$.  This improves the fit to 12.6-keV data from a $\chi^2/{\rm n.d.f.}$ of 2.16 to 1.25 (\fref{fig:Pix31Fit_Field_Diffusion}). 

   The diffusion mechanism has a greater effect on lower-energy spectra because energy deposited in the dead layer increases with the decreasing energy of the incident electron.  This effect is further enhanced by the influence of the magnetic field.  Backscattered electrons are returned to the detector and may make multiple transits of the dead layer.  The probability of backscattering also increases with decreasing electron energy.   
\begin{figure}[h] 
     \centering  
        \includegraphics[width=4.5in]{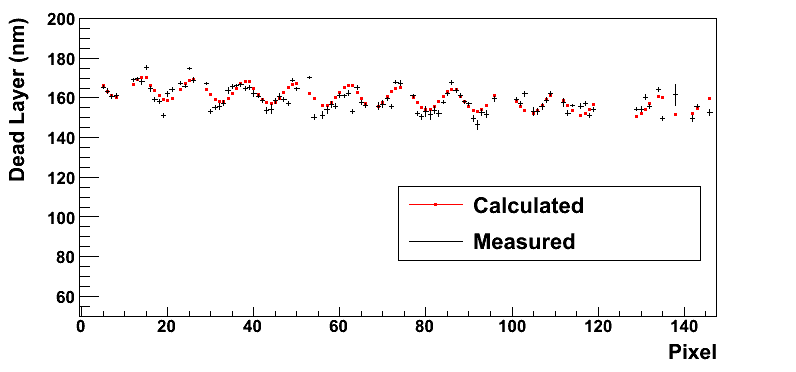} 
     \caption[Measured and Multi-Parameter Fit Calculated Dead-Layer Thickness]{Measured dead-layer thickness (black) and calculated dead-layer thickness (red) from the fit of Eq.~\ref{eq:globalDLequation} and parameters shown in the last line of \tref{Tab:globalFitParamters}, plotted against pixel number (see \fref{fig:PixelNumber}).}
     \label{fig:GlobalFitDLvsPixelNumber}
 \end{figure}
              
  Figure \ref{fig:GlobalFitDLvsPixelNumber} shows the dead-layer thickness (black), as determined from the $\chi^2$ minimization, for each active pixel with a charge-collection fraction of 0.46.  Two trends are seen in the plot: a 12-pixel periodic structure corresponding to the 12-pixel FPD rings, and a decrease in dead-layer thickness as the pixel number, and therefore the radial position, increases.  These geometric correlations are manifest in this plot because the pixel numbering scheme (\fref{fig:PixelNumber}) starts at a center pixel and spirals outward.  
  
  A  spurious periodic structure is  induced  by a rate variation of 200 to 3000\,Hz in electron-source data.  The UV light does not uniformly illuminate the electron-source disc, resulting in a nonuniform data rate across the focal-plane detector.  The data-acquisition system and electronics lack  baseline correction, causing rate-correlated effects in the output from the trapezoidal filters.  An increasing rate causes a deficit in the reconstructed energy and a degradation of the resolution.  The energy deficit is corrected by a free offset parameter in the spectral fits.  However, the rate dependence of the resolution remains.
  
   A simultaneous fit of rate ($s_i$), radial ($r_i$) and azimuthal ($\theta_i$) dependencies for each pixel is performed to include their effects on the determined values of the dead layer.  Assuming a nominal dead-layer thickness $D_N$ and that the dependencies are first- and second-order corrections, then a pixel's measured dead layer, $z_i$, can be described as:
 \begin{eqnarray}
z_i&=& D_N + Ar_i^2 + Gs_i+ B\cos(\theta_i)+ C\sin(\theta_i),
\label{eq:globalDLequation}
\end{eqnarray}
where $A$, $B$, $C$, and $G$ are additional parameters of the fit. The radial-dependence term $Ar_i^2$ is quadratic because the derivative must be zero at the origin. The two angular-dependence terms are required to produce an amplitude and phase in the azimuthal dependence.  The rate dependence of the data was found to be best described by a linear term $Gs_i$.
\begin{table}[h]
    \centering
    \caption[Multi-Parameter Dead-Layer Fit]{ Linear least-square fit parameters from Eq. \ref{eq:globalDLequation} to the FPD pixel dead-layer data.  The first row is the constant fit to the dead-layer data for $d=0.0$.  In each of the following rows a fit parameter is added to the minimization for the $d=0.46$ data.   }
\begin{tabular}{cccccccc}
\hline
$d$ & $\chi^2$ & n.d.f.  & $D_N$ (nm)  & $G$ (nm/Hz)  & $A$ (nm/cm$^2$) & $B$ (nm)& $C$ (nm)\\
\hline
\hline
0.0 & 12016 & 102  & $112.3 \pm 0.1$ &$\equiv 0$ & $\equiv 0$ & $\equiv 0$& $\equiv 0$ \\
0.46 & 3913& 101  & $161.5 \pm 0.1$ &$\equiv 0$ & $\equiv 0$ & $\equiv 0$& $\equiv 0$ \\
0.46 & 2281 & 100 & $155.4 \pm 0.2$ & $0.60 \pm 0.01 $& $\equiv 0$ & $\equiv 0$& $\equiv 0$ \\
0.46 & 1535 & 98  &$165.9 \pm 0.2$ & $\equiv 0$& $-0.59 \pm 0.02$ & $4.4 \pm 0.2 $& $-2.9 \pm 0.1$ \\
0.46 & 1423 & 99  & $161.1 \pm 0.3 $& $0.52 \pm 0.02 $& $-0.56 \pm 0.02 $&  $\equiv 0$& $\equiv 0$ \\
0.46 & 1193 & 97  & $160.9 \pm 0.3 $&$ 0.52 \pm 0.03 $&$ -0.56 \pm 0.02 $&$ -0.4 \pm 0.3$ &$ -1.8 \pm 0.1$ \\
\hline
\end{tabular}
\label{Tab:globalFitParamters}
\end{table}

 \begin{table}[h]
    \centering
    \caption[Correlation Matrix for Multi-Variable Dead-Layer Fit]{ Correlation matrix for the linear least-square fit parameters from Eq. \ref{eq:globalDLequation} to the focal plane detector pixels' dead-layer values. }
\begin{tabular}{c|ccccc}
\hline
&$D_N$ &$A$  & $B$ & $C$ & $G$ \\
\hline
$D_N$ & 1.00 & -0.56 & 0.61 & -0.28 & -0.81 \\
$A$ & -0.56 & 1.00 & 0.03 & -0.02 & 0.08 \\
$B$ & 0.61 & 0.03 & 1.00 & -0.42 & -0.84 \\
$C$ & -0.28 & -0.02 & -0.42 & 1.00 & 0.41 \\
$G$ & -0.81 & 0.08 & -0.84 & 0.41 & 1.00 \\
\hline
\end{tabular}
\label{Tab:globalFitCorrelationMatrix}
\end{table}

	The results of the linear least-squares fit of Eq. \ref{eq:globalDLequation} to the pixel dead layers, given in \tref{Tab:globalFitParamters}, show that the variations in dead layer are not purely a result of rate dependence.  The radial dependence is independent of the rate and azimuthal components as evidenced by the nearly zero values in the correlation matrix (\tref{Tab:globalFitCorrelationMatrix}). There is an expected correlation  between the azimuthal and rate components because the rate is strongly correlated with pixel azimuthal position.  

	The exact details of the ion implantation technique used to construct the FPD are not available.  It would not be surprising if the ion beam were to have had a radial dependence in implantation energy and/or intensity, causing a radial dependence in the measured dead layer.  The azimuthal component is either another deposition-technique artifact or a rate component to the dead layer that is not accounted for by the fit.  With the rate-dependence correction (\tref{Tab:globalFitParamters}), each pixel's measured dead layer results in a mean dead layer for the array of $155.4 \pm 0.5 \pm 0.2$\,nm.  The remaining geometric components account for a variation of $10.8 \pm 0.7$\,nm FWHM in the FPD dead layer.    
	
	The charge-collection fraction $d$ represents the fraction of charge created in the dead layer that actually arrives in the active depletion layer and participates in pulse formation.  For the KATRIN FPD, $d= 0.46$ results in the best description of the data, quite close to the value 0.50 that would be expected if diffusion alone governed the process.  
	
	The traditional alpha-particle measurement technique normally used by manufacturers in measuring the thickness of a dead layer yields a single quantity, the energy $\Delta E$ lost in the dead layer, and is (for layers thin enough that straggling is small compared to the instrumental resolution \cite{PhysRevA.11.1286}) unable to distinguish between a dead layer that is relatively thin with no charge recovery or thicker with partial charge recovery.  The dead-layer thickness measured by the alpha method is typically reported with the zero-charge-recovery value, but the true thickness $z_{\rm true}$ may be substantially larger.   To correct for charge recovery, Eq.~\ref{eqn:stoppingpower}  can be rewritten as follows: 
\begin{eqnarray}
z_{\rm true} &=& \Delta E (1-d)^{-1}\left< \frac{dE}{dx} \right>^{-1} \label{eqn:stoppingpowerrev}.
\end{eqnarray}
Based on Eq.~\ref{eqn:stoppingpower}, the dead layer of the KATRIN detector would, measured by the alpha method, be about 84\,nm, approximately half its true value.   The measured charge-collection fraction $d$ reported here can be used to correct alpha-particle measurements.  Only one wafer has been measured, and the value for $d$ may differ from wafer to wafer, but the arguments for a value of about 0.5 are quite general for this detector technology.  No alpha-particle measurement was made for the KATRIN FPD because the detectors were supplied as unmounted wafers.   As can be seen from the first row of \tref{Tab:globalFitParamters}, setting $d=0$ gives a lower value for the dead-layer thickness $D_N$, with electron beams as well.
	

%% file: Conclusion_DeadL_4_29_13_arXiv.tex
\section{Conclusion}\label{sec:conclusion}

The KATRIN experiment makes use of a monolithic, multipixel silicon \pin{} diode array to count electrons that pass its integrating energy spectrometers.   The electron line shape is particularly sensitive to microscopic details of electron energy loss in thin silicon dead layers because backscattered electrons can make several transits through the dead layer due to the magnetic and electrostatic field configuration.  A quantitative understanding of the line shape is desirable to optimize the analysis of the neutrino-mass data.  

Standard methods of measuring dead layers are impractical in the KATRIN apparatus.  Furthermore, the dead layer must be monitored periodically without removing the detector for {\em ex-situ} measurements.  We have presented a new method that involves normally incident electrons of varying energies.  

Comparison of the measured spectra with a detailed Monte Carlo calculation revealed that the observed line shape is quite sensitive to the model for the dead layer.  A widely used inert-slab model of a layer from which no charge is collected does not agree with data.  Instead, a diffusion model in which approximately half the charge generated in the dead layer is collected gives good agreement.   The mean dead layer of the KATRIN FPD is found to be $155.4 \pm 0.5$\,nm with a variation of $10.8 \pm 0.7$\,nm over the total area, and the fraction of charge collected from the dead layer is 0.46. 

%% file: ack.tex
\section{Acknowledgments}

Support has been provided by the US Department of Energy under grant DE-FG02-97ER41020, the German Helmholtz Gemeinschaft, and the Bundesministerium f\"{u}r Bildung und Forschung.  We gratefully acknowledge the participation and contributions of the entire KATRIN collaboration.